\documentclass[aps,twocolumn,showpacs]{revtex4}
\usepackage{graphicx}

\begin{document}

\title{Scaling Invariance: A Gateway to Phase Transitions}

\author{Edson D.\ Leonel}

\address{Departamento de F\'isica, Univ. Estadual Paulista - Unesp,
Av. 24A, 1515 - 13.506-900 - Rio Claro, SP, Brazil}

\date{\today} \widetext

\pacs{05.45.-a, 05.45.Pq, 05.45.Tp}

\begin{abstract}
We explore the concept of scaling invariance in a type of dynamical systems that undergo a transition from order (regularity) to disorder (chaos). The systems are described by a two-dimensional, nonlinear mapping that preserves the area in the phase space. The key variables are the action and the angle, as usual from Hamiltonian systems. The transition is influenced by a control parameter giving the form of the order parameter. We observe a scaling invariance in the average squared action within the chaotic region, providing evidence that this change from regularity (integrability) to chaos (non-integrability) is akin to a second-order or continuous phase transition. As the order parameter approaches zero, its response against the variation of the control parameter (susceptibility) becomes increasingly pronounced (indeed diverging), resembling a phase transition. These findings could not be obtained without a seminal paper on Phys. Rev. Lett. {\bf 2004}, {\em 93}, 014101.
\end{abstract}

\maketitle

\section{An Encounter}

My first contact with Professor Peter Vaughan Elsmere McClintock - whom I will affectionately refer to as Peter - occurred at the end of the year 2002 via email when I was approaching the completion of my Ph.D. in Physics. I was seeking a postdoctoral position, and Peter kindly welcomed me into his research group at Lancaster. I prepared a research proposal, applied for funding from the Brazilian agency CNPq, and was awarded a 12-month grant. After defending my Ph.D. in June 2003, I embarked on my journey to Lancaster at the end of July 2003. Initially, my stay was planned for one year. Still, the success of our results and the depth of the formalism we developed guided me to an extension of six months, allowing me to remain in Lancaster for one and a half years, hence coming back to Brazil in February 2005.

\begin{figure}[htb]
\centerline{\includegraphics[width=0.6\linewidth]{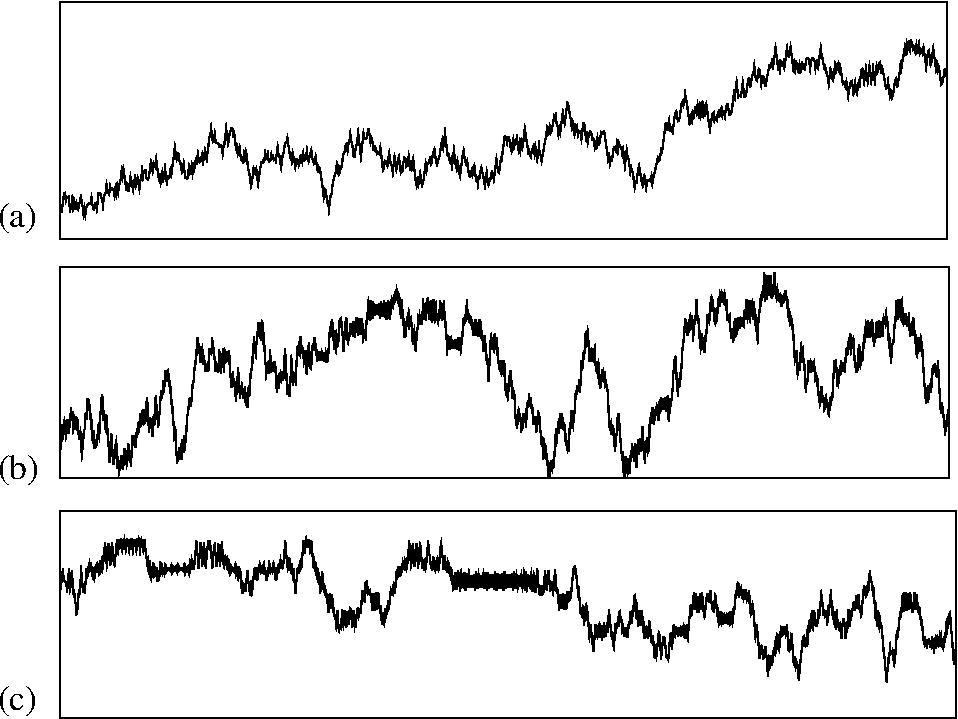}}
\caption{Plot of three chaotic series showing a regime of growth and decay evidencing similar characteristics for different time scales.}
\label{fig1}
\end{figure}

\begin{figure}[htb]
\centerline{\includegraphics[width=0.8\linewidth]{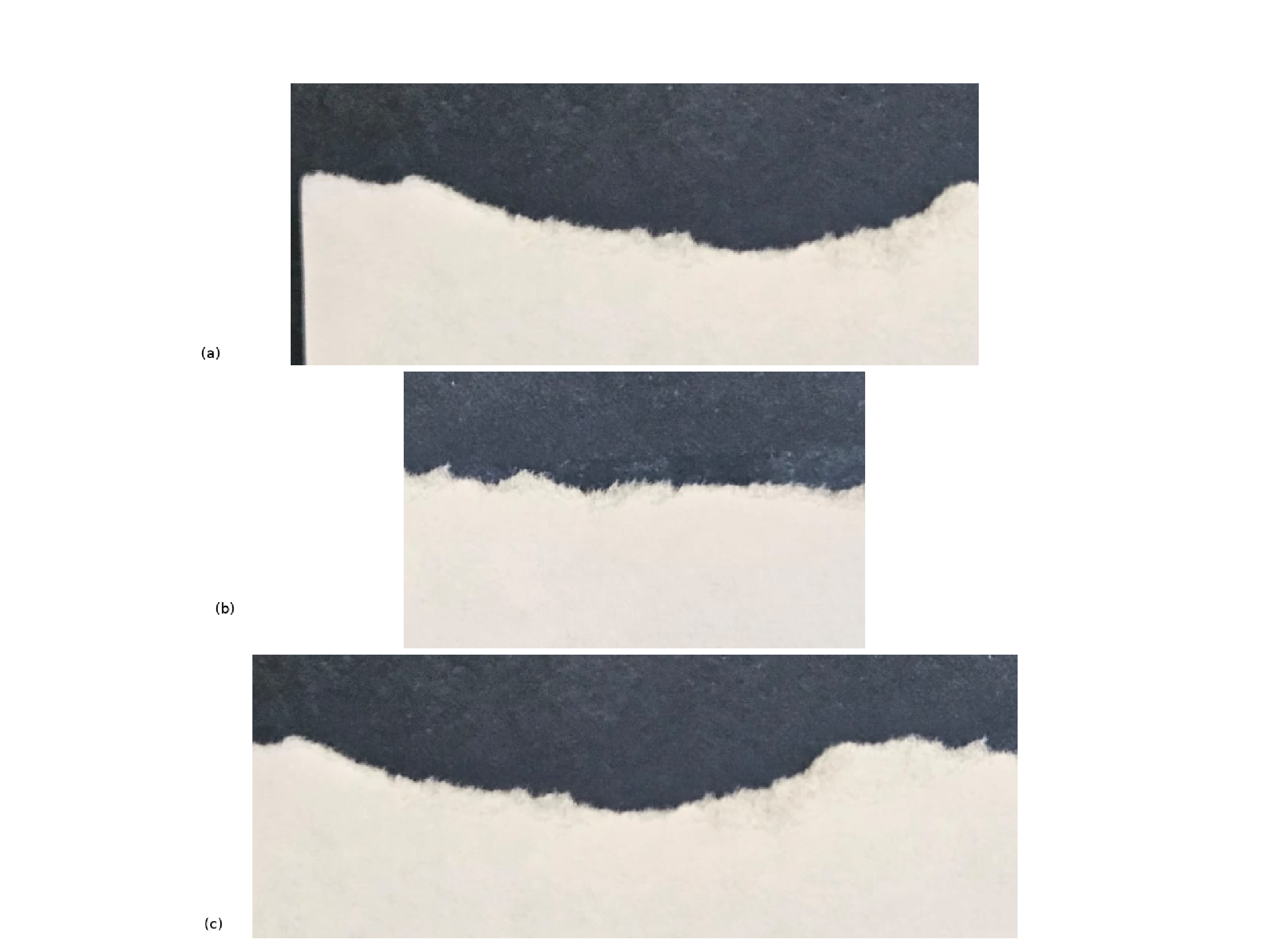}}
\caption{Figure showing three different pieces of scratched paper. The paper used was white A4, and the photo was taken against a dark blue background using a Samsung S24 cell phone camera.}
\label{fig2}
\end{figure}

Our project resulted in several significant publications, laying the foundation for my research trajectory. The insights we gained continue to influence ongoing projects and will undoubtedly shape future endeavors. We aimed to understand how chaotic diffusion could be characterized through scaling formalism. Towards the end of my Ph.D., I had observed intriguing similarities between chaotic orbits in Hamiltonian systems - when plotted as a function of time as shown in Fig. \ref{fig1} -
and the profiles of scratched paper, as illustrated in Fig. \ref{fig2}. 
It was already known \cite{barabasi} that profiles like those shown in Fig. \ref{fig2} in particle deposition lead to surface growth obeying scaling properties, which yields the definition of universality classes. The striking resemblance between the peaks and valleys of both scratched surfaces and chaotic trajectories, exhibiting a continuous interplay of growth and decay, suggests an underlying scaling invariance - an aspect that made our project both intellectually challenging and profoundly exciting.

The connection between scaling properties in chaotic orbits and universality classes became the central theme of my research during my time at Lancaster. The success of this work was reflected in several high-impact publications \cite{edl1,edl2,edl3,edl4,edl5,edl6,edl7}, most notably one in Physical Review Letters \cite{edl_prl}, which I shall revisit briefly throughout this paper.

During my time at Lancaster, Peter served as Head of the Physics Department, keeping him exceptionally busy. However, despite his many responsibilities, he always found time for noteworthy discussions and engaging interactions with the group. I fondly recall the pleasant lunches we shared, where his company was always a source of warmth and inspiration. It was gratifying to accept his invitation for a dinner shared with his wife, Marion McClintock, to whom I had the opportunity to talk about some historical points of Lancaster University. At that time, I had the privilege to know more about her work \cite{marion} and a little about the beginning of Lancaster University.

Peter's generosity, professionalism, and gentlemanly nature have left an enduring mark on me, shaping how I conduct myself as a researcher, mentor, and member of my academic community. The invaluable experiences and knowledge I gained under his guidance remain deeply treasured. For this, I am profoundly grateful.

One of Peter's passions in recent years has been hiking. During my last visit to him and Aneta Stefanovska in January 2024, they invited me on a wonderful hike in North Yorkshire, just after Storm Isha had swept across the UK with fierce, near-horizontal rain.
\begin{figure}[htb]
\centerline{\includegraphics[width=0.85\linewidth]{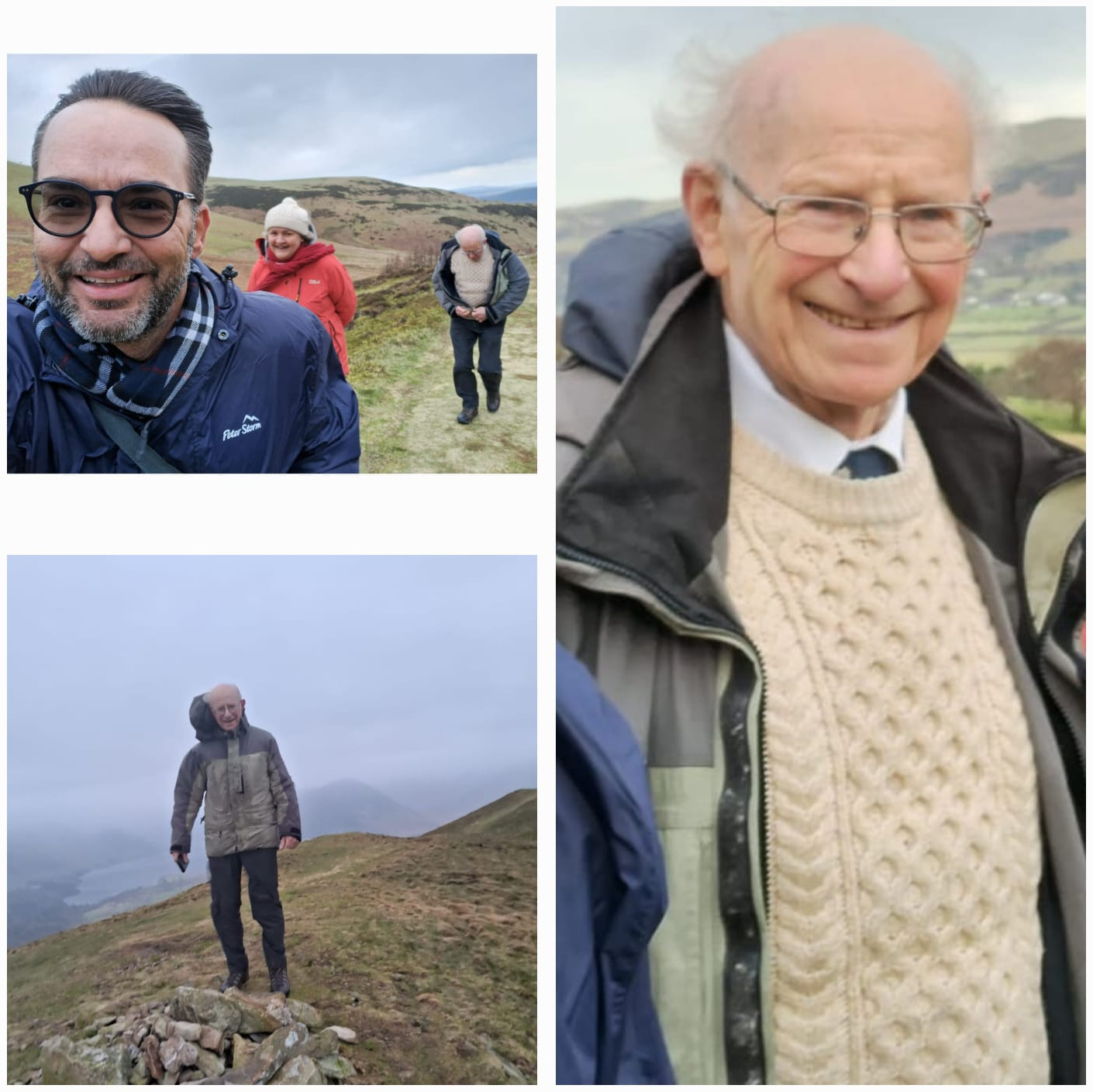}}
\caption{The top left shows Aneta, Peter, and me starting the hiking still under the effects of Isha's wind, making Peter fasten his jacket. The bottom left depicts Peter at the summit, bracing himself against the wind gusts. Finally, the right image beautifully captures Peter's joy and satisfaction at the successful completion of our hike.}
\label{fig3}
\end{figure}
Figure \ref{fig3} captures some memorable moments from our hike in North Yorkshire. The top left image shows Aneta, Peter, and me at the start of our journey, with the lingering effects of Isha's wind prompting Peter to fasten his jacket and increasing the size of my forehead by pulling my hair backward. The bottom left image depicts Peter at the summit, bracing himself against the powerful wind gusts. Finally, the image on the right beautifully captures Peter's joy and satisfaction with the successful completion of our hike.

As Peter celebrates his 85th birthday, I extend my warmest wishes for many more years of good health, happiness, and brilliance in physics. His contributions to the scientific community are indelible and remain a beacon of inspiration.

\section{Introduction}

As discussed in the previous section, the initial motivation for collaborating with Peter's group was to investigate scaling properties in chaotic orbits. Our approach began with dynamical systems described by mappings, focusing initially on a classic model known as the Fermi-Ulam model \cite{lich}.

The origins of this model trace back to Enrico Fermi's groundbreaking work in 1949 \cite{fermi}, where he sought to explain the exceptionally high energy of cosmic rays. Fermi hypothesized cosmic rays interact with interstellar space's fluctuating electric and magnetic fields. Given that these fields are time-dependent, such interactions could lead to energy gain.

\begin{figure}[htb]
\centerline{\includegraphics[width=0.5\linewidth]{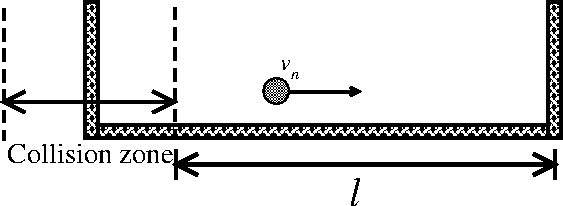}}
\caption{Plot of a schematic representation of the model.}
\label{fig4}
\end{figure}

Building upon Fermi's idea, Stanislaw Ulam formulated a mechanical model to illustrate this phenomenon. The system consists of a classical particle of mass $m$ confined between two rigid walls -- one fixed and the other oscillating over time. If the wall's motion is sufficiently smooth (i.e., possessing more than three continuous derivatives \cite{lich}), the particle does not experience unbounded energy growth. Figure \ref{fig4} shows a schematic representation of this setup.

At each collision with the moving wall, the particle's momentum may change, resulting in either an energy gain or loss depending on the phase of the wall at the moment of impact. The fixed wall merely redirects the particle back toward the moving boundary. The analogy to Fermi's original concept is evident: (i) {The particle represents the cosmic ray;} (ii) {The moving wall mimics the interactions with time-dependent electric and magnetic fields;} (iii) {The fixed wall acts as a mechanism to reinject the particle, analogous to repeated interactions with interstellar fields over time.}

\begin{figure}[t]
\centerline{\includegraphics[width=0.9\linewidth]{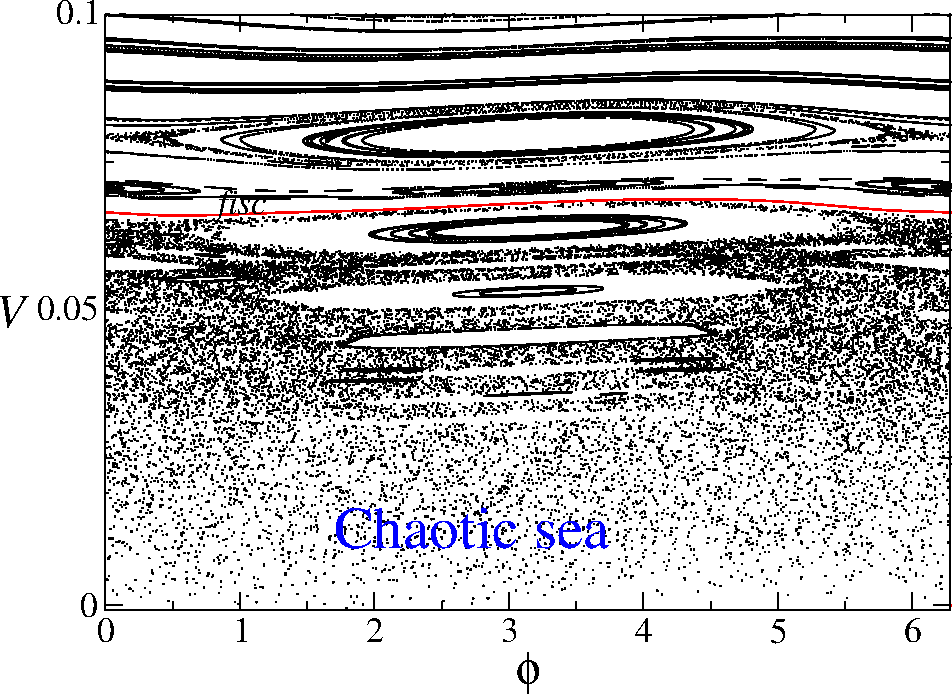}} \caption{Phase space representation of the Fermi-Ulam model. The axes correspond to the particle velocity $V$ and the phase of the moving wall $\phi$.}
\label{fig5}
\end{figure}

Ulam's formulation translates into a two-dimensional mapping, where the variables -- particle velocity $V$ and wall phase $\phi$ -- are updated at each collision with the moving boundary. If the collisions are purely elastic, meaning no energy is lost, the system's phase space divides into three distinct regions, as illustrated in Figure \ref{fig5}.

These regions are characterized as follows:
\begin{enumerate}
\item{Chaotic Region -- At low energy levels (comparable to the maximum kinetic energy of the oscillating wall), the system exhibits chaotic dynamics;}
\item{Stable Islands -- Periodic regions emerge, forming islands of stability where the motion remains regular;}
\item{Invariant Spanning Curves -- These structures act as barriers, preventing particles from transitioning between different energy regions.}
\end{enumerate}

A particularly significant feature of the phase space is the first invariant spanning curve (FISC), highlighted in Figure \ref{fig5}. The presence of such curve is crucial in determining the extent of the chaotic region, and we shall be back to it later in the paper. More importantly, it prevents the occurrence of Fermi acceleration, the phenomenon of unlimited energy growth in particles undergoing repeated collisions with moving boundaries. Since the FISC restricts particle movement across energy levels, unbounded acceleration is effectively suppressed in the system.

\section{The non-dissipative Fermi-Ulam model}

In this section, we characterize the average velocity and its variance within the chaotic sea of the phase space, which we made using a scaling approach. The formalism was used, so far we can say, for the first time in the seminal publication signed by us in Physical Review Letters \cite{edl_prl}. It provides a characterization of the integrability-to-chaos transition in the Fermi-Ulam model. We consider a classical particle bouncing between two rigid walls to describe the system. One is fixed, and the other is moving periodically in time with a normalized amplitude $\epsilon$. The system is modeled using the mapping $T(V_n,\phi_n)=(V_{n+1},\phi_{n+1})$ which determines the velocity of the particle and the phase of the time-moving wall immediately after a collision. Our investigations were made using the static wall approximation \cite{lich}. Instead of explicitly moving the wall, we assume both walls remain fixed, but when the particle collides with one of them (say, the one on the left), it exchanges momentum as if the wall were in motion. This simplification accelerates numerical simulations while preserving the system's key dynamical properties.

With this simplification and using dimensionless variables, the mapping is given by \cite{lich}
\begin{equation}
T:\left\{\begin{array}{ll}
V_{n+1}=|V_n-2\epsilon\sin(\phi_{n+1})|~~\\
\phi_{n+1}=[\phi_n+{2\over{V_n}}]~ {\rm mod} 2\pi\\
\end{array}
\right..
\label{eq1}
\end{equation}

The time interval during the flight between the collisions is given by  $2/V_n$ while $-2\epsilon\sin(\phi_{n+1})$ corresponds to the velocity gained or lost in the collision. The modulus function is introduced to ensure that the particle remains confined within the region between the walls. We stress that the approximation of using the simplified FUM is valid in the limit of small $\epsilon$. Therefore, the transition from integrability ($\epsilon=0$) to chaos ($\epsilon\ne 0$), characterizing the birth of the chaotic sea, can be well described.

We concentrated on the scaling behavior present in the chaotic sea. We investigate the evolution of the velocity averaged in $M$ initial phases, namely
\begin{equation}
V(n,\epsilon,V_0)={1\over M}\sum_{j=1}^M V_{n,j}~~,
\label{eq_jaff_1}
\end{equation}
where $V_0$ is the initial velocity and $j$ refers to a ensemble sample.

To define the deviation around the average velocity, we first consider the velocity averaged over the orbit generated from a single initial phase
\begin{equation}
\overline{V}(n,\epsilon,V_0)={1\over{n}}\sum_{i=0}^n V_i~.
\end{equation}
We then evaluate the interface width around this averaged velocity. Finally, the deviation around the average velocity is given considering an ensemble of $M$ different initial phases:
\begin{equation}
\omega(n,\epsilon,V_0) \equiv \frac{1}{M} \sum_{j=1}^{M}
\left[ \sqrt{\overline{V^2}_j(n,\epsilon,V_0) - \overline{V}_j^2(n,\epsilon,V_0)}~\right].
\end{equation}

\begin{figure}[t]
\centerline{\includegraphics[width=0.7\linewidth]{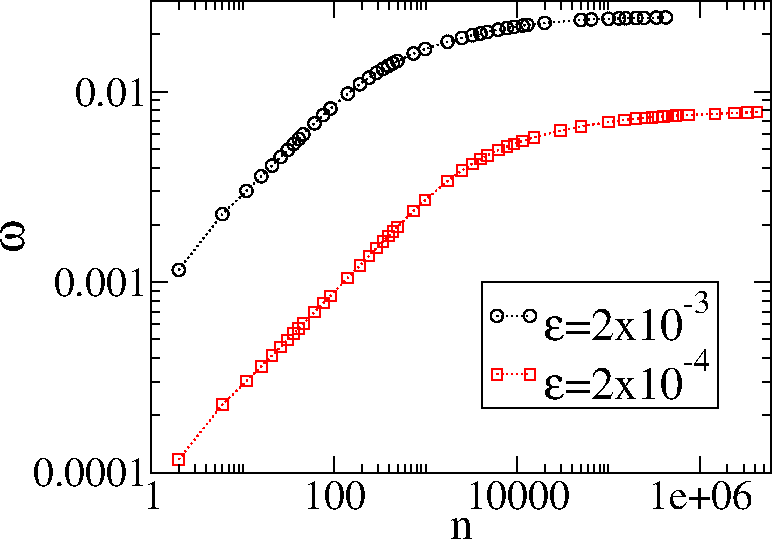}}
\caption{Behaviour of $\omega~vs.~n$. The curves were generated from an ensemble average of $5\times 10^4$ different initial conditions starting with $V_0\approx 0$.}
\label{fig6}
\end{figure}

\begin{figure}[t]
\centerline{\includegraphics[width=0.7\linewidth]{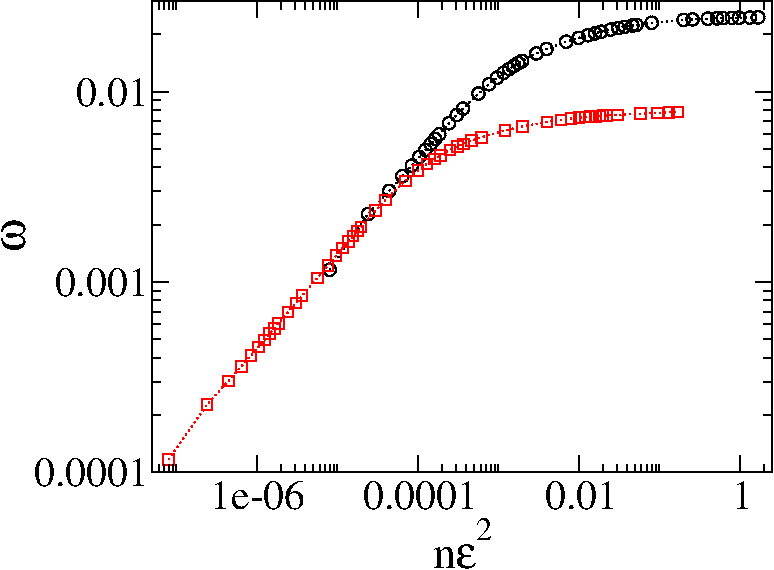}}
\caption{Behaviour of $\omega~vs.~n\epsilon^2$ showing the two curves merging their initial behavior of growth.}
\label{fig1_a}
\end{figure}

Figure \ref{fig6} illustrates the behavior of $\omega$ for two different control parameters.
From Fig. \ref{fig6}, we see that $\omega$ grows for small $n$, passes from a crossover, and approaches a regime of saturation for large $n$. The change from growth to saturation is characterized by a crossover $n_x$. We also notice that different control parameters generate different curves that have similar behavior but occupy different places in the figure. However, an ad-hoc transformation $n\rightarrow n\epsilon^2$ coalesces all curves to start to grow together for short $n$. Figure \ref{fig1_a} shows the two curves plotted in Fig. \ref{fig6} after the transformation $n\rightarrow\epsilon^2$.

As we have seen from Figs. \ref{fig6} and \ref{fig1_a} generated from different control parameters, the curves start to grow as a power law in $n$, and after passing from a crossover, they bend towards a regime of saturation. Moreover, the behavior is similar despite the control parameter. This typical behavior observed in scaling invariance can be described using the scaling approach. We therefore suppose that: (i) for $n \ll n_x$, $\omega$ grows as
\begin{equation}
\omega(n\epsilon^2,\epsilon,V_0)\propto ({n\epsilon^2})^{\beta}~,
\label{eq5}
\end{equation}
where $\beta$ is the acceleration exponent; (ii) for $n \gg n_x$, $\omega$ reaches a saturation regime given as
\begin{equation}
\omega_{\rm sat}(\epsilon)\propto \epsilon^{\alpha}~, \label{eq6}
\end{equation}
where $\alpha$ is the saturation exponent; and (iii) the crossover iteration number $n_x$ giving the changeover to the saturation is
\begin{equation}
n_x(\epsilon,V_0)\propto\epsilon^z~,
\label{eq7}
\end{equation}
where $z$ is a dynamical exponent. These hypotheses allow us to describe $\omega$ as a scaling function of the type
\begin{equation}
\omega(n\epsilon^2,\epsilon,V_0)=l\omega(l^an\epsilon^2,l^b\epsilon,l^cV_0)~,
\label{eq8}
\end{equation}
where $l$ is the scaling factor, $a$, $b$, and $c$ characteristic exponents. 

\begin{figure}[b]
\centerline{\includegraphics[width=1.0\linewidth]{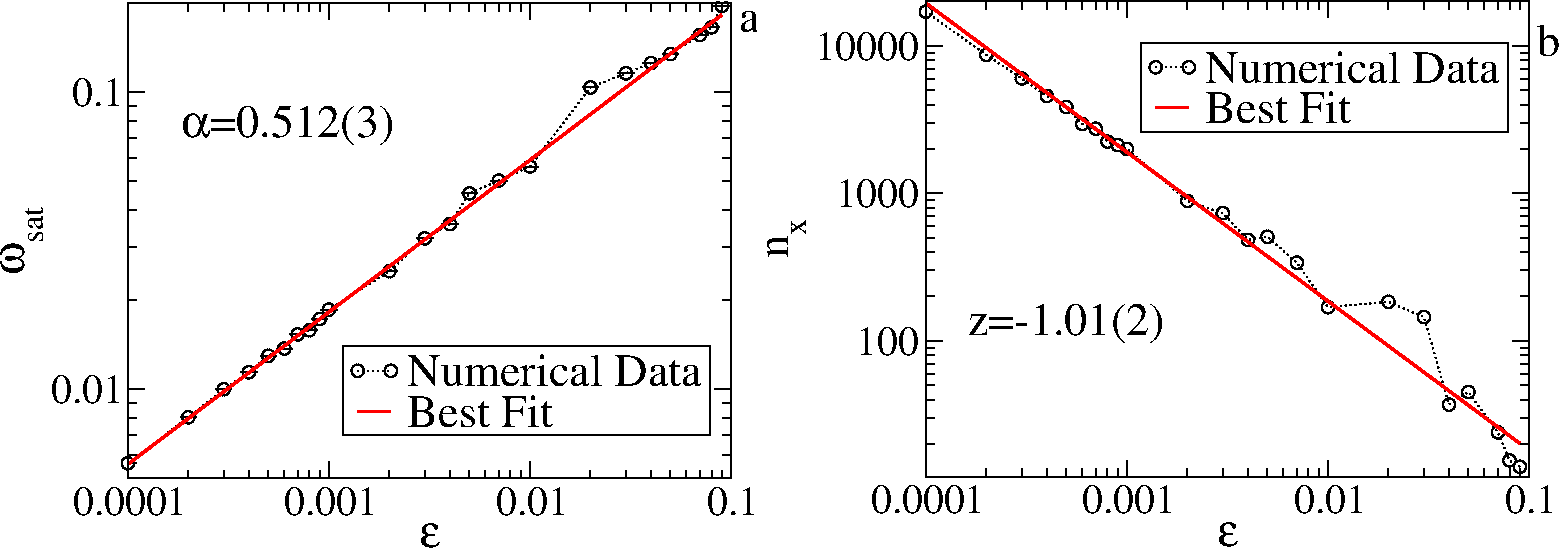}}
\caption{(a) Plot of $\omega_{\rm sat}~vs.~\epsilon$. (b) The crossover iteration number $n_x~vs.~\epsilon$. }
\label{fig8}
\end{figure}
\begin{figure}[t]
\centerline{\includegraphics[width=1.0\linewidth]{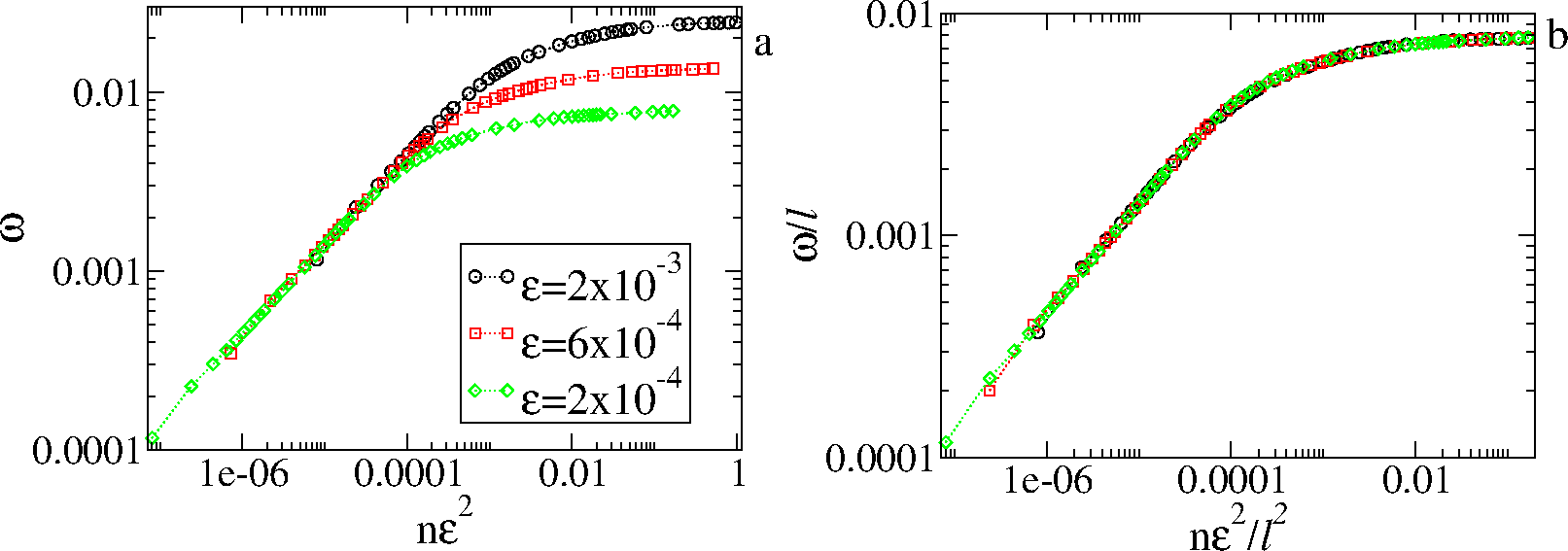}}
\caption{(a) Plot of $\omega$ for different $\epsilon$. (b) Overlap of the curves from (a) onto a universal curve. Both (a) and (b) were obtained using $V_0\approx 0$.} \label{fig9}
\end{figure}
\begin{figure}[b]
\centerline{\includegraphics[width=1.0\linewidth]{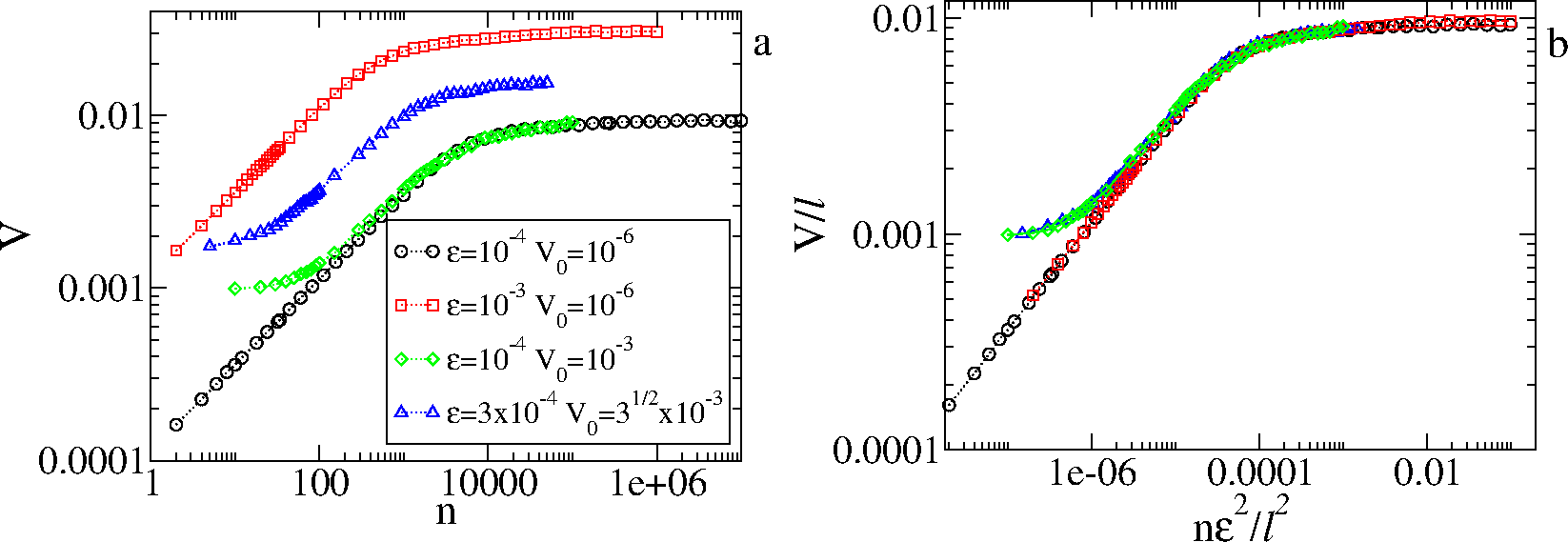}}
\caption{Behaviour of the average velocity $V~vs.~n$ for different values of $\epsilon$ and $V_0$. (a) The original time series. (b) Overlap of curves into a single and universal curve.}
\label{fig10}
\end{figure}

We start by chosing $l=(n\epsilon^2)^{-{1\over{a}}}$. This allows us to rewrite (\ref{eq8}) as
\begin{equation}
\omega(n\epsilon^2,\epsilon,V_0)=(n\epsilon^2)^{-{1\over{a}}}
\omega_1\left((n\epsilon^2)^{-{b\over{a}}}\epsilon,
(n\epsilon^2)^{-{c\over{a}}}V_0\right)~.
\label{eq9}
\end{equation}
The function
$\omega_1=\omega\left(1,(n\epsilon^2)^{-{b\over{a}}}\epsilon,
(n\epsilon^2)^{-{c\over{a}}}V_0\right)$ is assumed to be a constant for
$n \ll n_x$. Comparing equations (\ref{eq9}) and (\ref{eq5}), obtain $-{1\over{a}}=\beta$. Choosing now $l=\epsilon^{-{1\over{b}}}$, we find
\begin{equation}
\omega(n\epsilon^2,\epsilon,V_0)=\epsilon^{-{1\over{b}}}
\omega_2\left((n\epsilon^2)\epsilon^{-{a\over{b}}},
\epsilon^{-{c\over{b}}}V_0\right)~,
\label{eq10}
\end{equation}
where $\omega_2=\omega\left(
(n\epsilon^2)\epsilon^{-{a\over{b}}},1,
\epsilon^{-{c\over{b}}}V_0\right)$ is assumed constant for $n \gg
n_x$. A comparison of Eqs. (\ref{eq10}) and (\ref{eq6}) shows
that $-{1\over{b}}=\alpha$. 

Let us now discuss how to obtain $c$. We shall use a connection with an important transition in the Chirikov-Taylor map \cite{lich} to do that. For small control parameters, the phase space for the Chirikov-Taylor map shows a mixed form with periodic islands, chaotic sea, and invariant spanning curves (invariant tori). As soon as the control parameter increases, the phase space changes, chaotic sea increases, and the invariant spanning curves are reduced, lasting only those stable whose returning times are obtained along those in the Fibonacci sequence \cite{joelson1,joelson2} and obey the Slater Criteria \cite{celso}. The last one to be destroyed happens at the control parameter $K_c=0.9716\ldots$. All invariant spanning curves are destroyed for any control parameter larger than $K_c$, allowing the chaotic orbits to diffuse unbounded in the phase space. In contrast, the invariant spanning curves prevent the unbounded diffusion for $K<K_c$. Therefore, at $K_c$ there is a transition from local $K<K_c$ to globally chaotic behavior $K>K_c$. In the FUM, below the first invariant spanning curve, there is only a large chaotic sea surrounding periodic orbits, while above it, there might be periodic and some local chaotic dynamics. The first invariant spanning curve limiting the size of the chaotic domain in the FUM can be described as a local approximation of the Chirikov-Taylor map. Using the appropriate mathematical procedure as made in Ref. \cite{joelson1,joelson2}, the position of the first invariant spanning curve (FISC) placed above the chaotic sea has a control parameter $\epsilon$ linked with a typical mean velocity along the FISC and give an effective control parameter $K_{\rm
eff}=4\epsilon/{V^*}^2\approx 0.9716\ldots$. We rewrite the effective control parameter $K_{\rm eff}$ in terms of scaled variables as
\begin{equation}
K_{\rm eff}={4(l^b\epsilon)\over{(l^cV_0)^2}}=
{4\epsilon\over{V_0^2}}{l^b\over{l^{2c}}}~. \label{eq11}
\end{equation}
leading to $b-2c=0$. Our result for the exponent $b$ gives $c=-{1\over{2\alpha}}$. All characteristic exponents are determined if $\alpha$ and $\beta$ are known.

The asymptotic state is obtained at the limit of large $n$, which is the limit to obtain the exponent $\alpha$. It is also independent of $V_0$. Figure \ref{fig8}(a) shows a plot of $\omega_{sat}~vs.~\epsilon$. A power law fit gives $\alpha=0.512(3)\cong 1/2$. Equation (\ref{eq8}) is rewritten as
\begin{equation}
\omega(n\epsilon^2,\epsilon,V_0)=
(n\epsilon^2)^{\beta}g\left[(n\epsilon^2)^{-2\beta}\epsilon,
(n\epsilon^2)^{-\beta}V_0\right]~.
\label{eq12}
\end{equation}

The acceleration exponent is obtained at the limit of $V_0\cong 0$ before saturation. For a range of $\epsilon\in [10^{-4},10^{-1}]$ we obtain $\beta=0.496(6)\approx 1/2$ leading to $a=b=-2$ and
$c=-1$. Therefore, from the Eqs. (\ref{eq7}) and (\ref{eq9}) we obtain a scaling law $z=\alpha/\beta-2$. It is straightforward to obtain for the FUM that $z=-1$. Figure \ref{fig8}(b) shows the behavior of $n_x$ as a function of the control parameter $\epsilon$. A power law fit gives us that
$z=-1.01(2)$, in good agreement with the scaling result. The scaling
for $V_0\approx 0$ is shown in Fig. \ref{fig9}, where the
three different curves for $\omega$ in (a) are overlapped onto a single and hence universal curve seen in (b).

The average velocity better illustrates the additional crossover that depends on the initial velocity. We consider two ``time'' scales, namely $n_x^{\prime}\propto 1/\epsilon$ and $n_x^{\prime\prime}\propto V_0^2/\epsilon^2$. From  Eq. (\ref{eq11}), the larger initial velocity in the chaotic sea scales with $V_{0,{\rm max}}\approx 2\epsilon^{1/2}$ implying that the second time scale has a maximum value of ($n_x^{\prime\prime}\sim 4n_x^{\prime}$). We notice two different kinds of behavior for $n_x^{\prime\prime}<n_x^{\prime}$ or $n_x^{\prime\prime}\sim
n_x^{\prime}$. When $V_0=10^{-6}$, we have $n_x^{\prime\prime}\approx 0$. From Fig. \ref{fig10}(a) we see the curves for $\epsilon=10^{-4}$ and $\epsilon=10^{-3}$ show only two regimes: (1) a growth in power law for $n \ll n_x^{\prime}$ and (2) the saturation regime for $n \gg n_x^{\prime}$. Considering $V_0=10^{-3}$ and
$\epsilon=10^{-4}$ we have that $n_x^{\prime\prime}<n_x^{\prime}$
and we see three regimes. For $n\ll n_x^{\prime\prime}$, the average velocity is constant. When $n_x^{\prime\prime}<n<n_x^{\prime}$, the
curve growth and begin to follow the curve of $V_0=10^{-6}$ and
same $\epsilon$. In this range of $n$, we have a growth with a
smaller effective exponent $\beta$. Finally, we have the saturation regime for $n\gg n_x^{\prime}$. It is shown in Fig. \ref{fig10}(b) that the overlap of the curves holds even for
$V_0\not=0$, implying that the inferred scaling form
$V(n\epsilon^2,\epsilon,V_0)$ with exponents $a=b=-2$ and $c=-1$
is also correct.

\section{Critical exponents}

As discussed in the previous section, diffusion within the chaotic sea is inherently limited. When an initial condition is set in the chaotic region, the orbit evolves within this domain, exhibiting fluctuations in velocity - both increasing and decreasing. However, there exists an upper bound to this growth. The first invariant spanning curve acts as a barrier, preventing chaotic orbits from crossing it and thereby setting an upper limit on the size of the chaotic domain.

As particles diffuse within the chaotic sea, the diffusion behavior is described by a power law for a low initial velocity. The curve of the average velocity and the deviation of the average velocity undergoes a changeover from the regime of growth, hence approaching a domain of constant plateau. The growth regime is marked by a power law whose critical exponent $\beta$ determines the acceleration of the growth. The changeover is defined by a critical exponent $z$, while the saturation of the curves obtained for a long enough time is marked by a critical exponent $\alpha$. The set of three critical exponents leads to a universal scaling law that establishes an analytical relation between the exponents, given by
\begin{equation}
z={{\alpha}\over{\beta}}-2.
\end{equation}

The numerical values of the critical exponents can be used to define universality classes. For the Fermi-Ulam model, the critical exponents are $\alpha=0.5$, $\beta=0.5$ and $z=-1$. Interestingly, these exponents can identify a phase transition from integrability to non-integrability observed when the control parameter changes from $\epsilon=0$ to any $\epsilon\ne 0$.

The periodically corrugated waveguide \cite{leonel_waveguide} is a rather different system that exhibits a transition from integrability to non-integrability yet belongs to the same universality class as the Fermi-Ulam model. The model consists of a classical light ray that is specularly reflected between a corrugated surface given by
\begin{equation}
y=y_0+d\cos(k x)
\end{equation}
and a flat plane surface at $y=0$. Here $y_0$ denotes the average distance between the corrugated and flat surfaces, $d$ is the corrugation amplitude, and $k$ is the wave number. The dynamical variables used in the problem description are the angle $\theta$ of the ray's trajectory measured from the positive horizontal axis and the corresponding value of the $x$ coordinate at the instant of reflection. Figure \ref{wave} illustrates the behavior of a light ray reflection:

\begin{figure}[t]
\centerline{\includegraphics[width=0.85\linewidth]{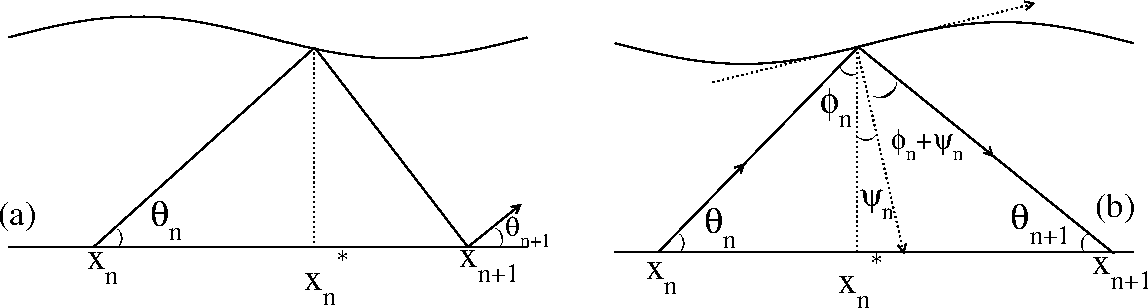}}
\caption{(a) Reflection from the corrugated surface of a light ray
coming from the flat surface at $y=0$. (b) Details of the
trajectory before and after a collision with the corrugated
surface.}
\label{wave}
\end{figure}

A two-dimensional, area-preserving mapping  describes the dynamics of the light ray \cite{leonel_waveguide}. The phase space is composed of three parts: (i) a sizeable chaotic sea surrounding (ii) periodic islands, and (iii) a set of invariant spanning curves, with the lowest one limiting the size of the chaotic sea and blocking the passage of particles through it.

Interestingly, the chaotic diffusion obeys remarkably similar conditions to those observed in the diffusion of the Fermi-Ulam model. A striking result is that the three critical exponents observed for the transition from integrability to non-integrability in the periodically corrugated waveguide are the same as those for the Fermi-Ulam model. Even though the models are fundamentally different, the observed transition identifies them as belonging to the same universality class.

\section{An extended model}

As we have seen by two models in the previous sections, there is a transition from integrability to non-integrability with specific critical exponents $\alpha$, $\beta$, and $z$. They lead to the knowledge of universality classes, which is typical of the scaling approach. The formalism defined during my visit to Peter's group primarily allows us to identify the type of phase transition the systems are undergoing. 

In this section, we explore the concept of scaling invariance in a dynamical system that transitions from order (regularity) to disorder (chaos), hence generalizing the earlier discussion from the previous sections. The systems we consider are described by a two-dimensional, nonlinear mapping that preserves the area in the phase space. The key variables are the action and the angle, as usual from Hamiltonian systems. The transition is influenced by a control parameter giving the form of the order parameter. We observe a scaling invariance in the average squared action within the chaotic region, providing evidence that this change from regularity (integrability) to chaos (non-integrability) is akin to a second-order or continuous phase transition. As the order parameter approaches zero, its response against the variation of the control parameter (susceptibility) becomes increasingly pronounced, resembling a phase transition.

To do the investigation and following some of our previous results \cite{pub1,pub2,pub3,pub4,book1,book2}, we consider a generic system described by two degrees of freedom whose dynamics are given by a Hamiltonian written as $H(I_1,\theta_1,I_2,\theta_2)=H_0(I_1,I_2)+\epsilon H_1(I_1,\theta_1,I_2,\theta_2)$. Here, $H_0$ represents the organized dynamics, maintaining the system's integrability, while $H_1$ adds a touch of nonlinearity, controlled by a parameter $\epsilon$.

Think of $\epsilon$ as tunning parameter. When turned null ($\epsilon=0$), the system preserves energy and action. However, when it is increased ($\epsilon\neq 0$), things may get wild as only energy is preserved. This shift, from well-organized dynamics to a more complex one, resembles a continuous phase transition, analogous as observed in statistical mechanics \cite{b2,b3,b4}. 

Since $H$ is time independent \cite{b5}, the symmetry leads the dynamics to be described by a three-dimensional flux that, when intercepted by a Poincar\'e section, gives a two-dimensional mapping of the type
\begin{equation}
\left\{\begin{array}{ll}
I_{n+1}=I_n+\epsilon h(\theta_n,I_{n+1})\\
\theta_{n+1}=[\theta_n+K(I_{n+1})+\epsilon p(\theta_n,I_{n+1})]~~{\rm mod
(2\pi)}\\
\end{array}
\right..
\label{eq0}
\end{equation}
Here, both $K(I_{n+1})$, $p(\theta_n,I_{n+1})$ and $h(\theta_n,I_{n+1})$ are continuous functions of their variables and $n$ gives the iterated of the mapping. It only preserves the area on the phase space if ${{\partial p(\theta_n,I_{n+1})}\over{\partial \theta_n}}+{{\partial h(\theta_n,I_{n+1})}\over{\partial I_{n+1}}}=0$ is attended.

We consider a specific family of systems described by
\begin{equation}
\left\{\begin{array}{ll}
I_{n+1}=I_n+\epsilon \sin(\theta_n)\\
\theta_{n+1}=[\theta_n+{{1}\over{|I_{n+1}|^{\gamma}}}]~~{\rm mod (2\pi)}\\
\end{array}
\right.,
\label{eq1_new}
\end{equation}
where $I$ and $\theta$ represent the dynamical variables of the system. The parameter $\epsilon$ influences the evolution of the system. As we mentioned earlier, the motivation behind this particular formulation is associated with chaotic diffusion. When the action $I$ is small, the variable $\theta_{n+1}$ is uncorrelated with $\theta_n$, leading to chaotic orbits and allowing the action to increase (diffuse in the phase space). As the action grows, angular variables correlate, introducing regularity in the phase space. This regularity manifests as periodic islands and invariant spanning curves (invariant tori), influencing the dynamics significantly.

Our primary objective is to explore diffusion in the phase space, which shows a rather crucial scaling invariance observed in a transition from integrability to non-integrability. We focus on answering four guiding questions - (1) Identify the broken symmetry: Pinpoint where the symmetry in the system is disrupted. (2) Propose an order parameter: it gives an observable related to the dynamical variables, which attends to the requirements that it goes to zero at the transition. In contrast, its susceptibility (response of the order parameter to the variation of the control parameter) diverges in the same limit. (3) Discuss the elementary excitation: Understand the influence of the elementary excitations of the system leading to a diffusive behavior. (4) Discuss the topological defects: Examine unexpected structural elements directly impacting particle transport.

To start with, let us delve into the symmetry of the problem.

\subsection{Broken symmetry}

\begin{figure}[t]
\centerline{\includegraphics[width=1.0\linewidth]{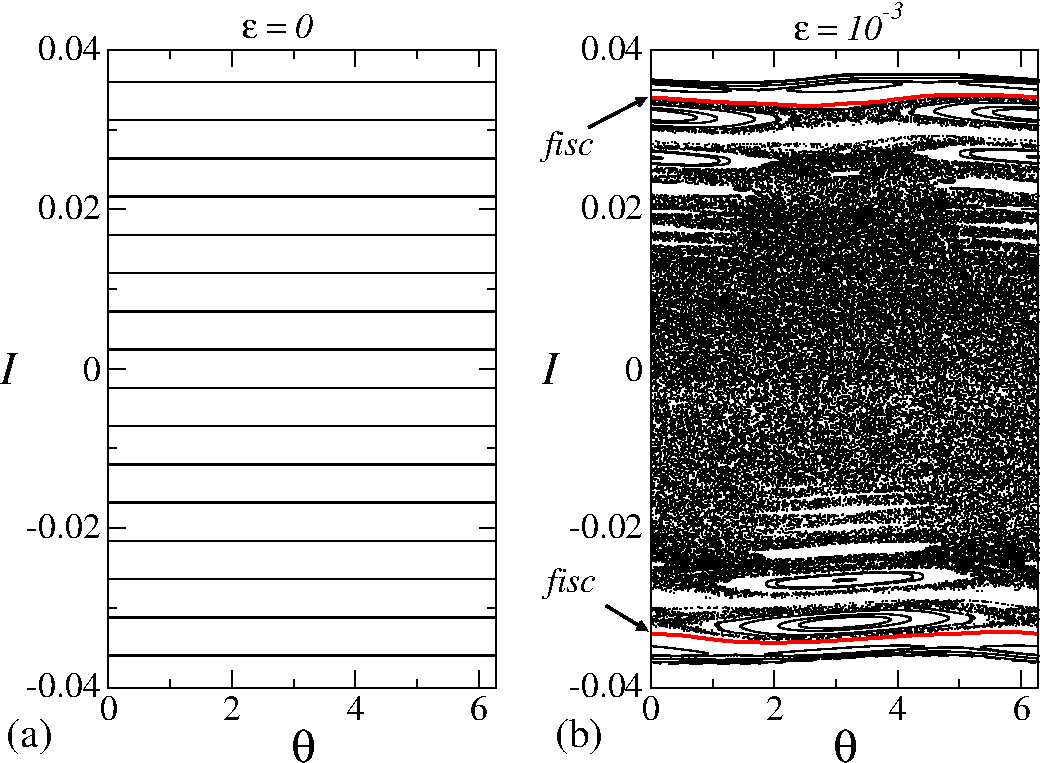}}
\caption{Plot of the phase space for the mapping (\ref{eq1_new}) 
for: (a) $\epsilon=0$ and (b) $\epsilon=10^{-3}$. The curves shown in (b) correspond to the first invariant spanning curves and scale with $\epsilon^{1/(1+\gamma)}$.}
\label{Fig1}
\end{figure}

We discuss some of the characteristics of the phase space and their influences on the dynamics. We note that the parameter $\epsilon$ plays a key role in shaping the dynamics. When $\epsilon=0$, the system is considered integrable because the energy and action remain constant. Picture this phase as a neat arrangement in space, marked by a phase space with a constant action, as shown in \ref{Fig1}(a). In this scenario, the system's behavior is entirely predictable, with no rapid (exponentially) spreading of nearby initial conditions. It's a kind of well-behaved phase.

However, things get far more interesting when $\epsilon \neq 0$. The once tidy phase space turns into a mix of somewhat different complexities. We witness a dynamic interplay depending on the starting conditions and the control parameters. The phase space now hosts a chaotic sea, surrounded by invariant spanning curves and dotted with periodic islands as shown in \ref{Fig1}(b).

Thanks to Liouville's theorem \cite{b5} and the preservation of area in the phase space, stability islands act like guardians. They keep particles within the chaotic sea from wandering off and prevent particles inside from escaping. It's almost like these stability islands are the phase space's version of topological defects \cite{b6}, disrupting the expected flow of particles and violating the usual predictability and, hence, the ergodicity. Figure \ref{Fig1} shows a plot of the phase space for the mapping (\ref{eq1_new}) considering the parameters: (a) $\epsilon=0$ and (b) $\epsilon=10^{-3}$.

The phase space also exhibits a set of curves traversing the entire configurational space, and they depend on the control parameter. Now, let us dive into the concept of spanning curves. These unique curves have a fascinating property -- they remain unchanged over time. If you start something along one of these curves, it keeps evolving along that curve indefinitely. It is like having a path in the system that, once you start on it, you are on a journey that lasts forever. These curves are crucial in shaping the system's long-term behavior and are pivotal in this dynamic. By blocking the movement of particles from one side to the other, these curves define the size of the chaotic sea. They act as barriers, shaping the space where chaos can freely unfold.

As we discussed earlier about the chaotic sea -- a zone of unpredictability in the phase space -- it is interesting to note that it has a specific length along the action axis. Think of the chaotic sea as a constantly changing space where, if you place an initial condition, it can stretch over a particular range of actions, from negative to positive values. However, here is where it gets fascinating. If you start with two initial conditions very close to each other in this chaotic sea, they drift apart exponentially over time. It is like watching the chaos unfold in an ever-expanding manner. However, the diffusion is always limited by the invariant spanning curves. They act like barriers, preventing particles from crossing through. So, if you are in the chaotic sea and hit one of these curves, you can not go any further. It's like having invisible boundaries that confine the chaos within a specific size, shaping the behavior of our system.

Now, let us go a crucial point that the previous discussion sets the stage for -- understanding the broken symmetry in the system when $ \epsilon=0$, the phase space has a unique and regular structure. Each curve in Figure \ref{Fig1}(a) depends solely on the initial action, which remains constant throughout the dynamics. Because the action is preserved, nearby initial conditions do not drift apart exponentially over time -- a key condition for chaotic dynamics. So, in this scenario, we witness a regular dance in the phase space, representing a phase of orderliness in the dynamics.

Contrastingly, when $\epsilon\ne0$, the nonlinear function $\sin(\theta)$ steps into the spotlight, influencing the time evolution of particles and disrupting the orderliness present in the phase space. Now, the phase space takes on a mixed form, featuring periodic dynamics with fixed points, invariant spanning curves (shown as continuous curves in Figure \ref{Fig1}(b)), and a chaotic sea. Within the chaotic sea, something interesting happens. Two nearby initial conditions exponentially drift apart over time, a characteristic feature of chaotic dynamics. This chaotic sea marks a phase of unpredictability and disorder in the system.

The shift from regularity to chaos, triggered when $\epsilon\ne0$, signifies symmetry breaking in the system. This change defines the window size for chaotic dynamics. In simpler terms, when $\epsilon$ takes on a non-zero value, the once orderly and predictable behavior gives way to chaos. This transition from regularity to chaotic dynamics, in turn, leads the particle to have chaotic diffusion.

Additionally, we observe that the mixed phase space, featuring islands and invariant spanning curves, shows different averages for chaotic diffusion. Whether we measure it across various initial conditions or over time, this difference holds true. This introduces a key distinction: the time average differs from the microcanonical average.

This deviation from uniform behavior breaks the fundamental assumption of ergodicity, meaning that the system does not exhibit the expected homogeneity. In simpler terms, different parts of the phase space do not mix freely. Chaotic dynamics, for instance, cannot infiltrate periodic structures, and vice versa.

Picture a particle navigating the chaos. As it cruises through the unpredictable dynamics, there is a fascinating twist. When it comes close to periodic structures or finds itself on an invariant spanning curve, it can get momentarily stuck -- this intriguing occurrence is called "stickiness" \cite{ref2}. This sticky situation has a notable effect: it alters the probability distribution of finding a particle with a particular action at a specific time.

Our earlier discussion emphasized the broken symmetry in the phase space. However, there is another equally crucial aspect to consider. If we focus on the first equation of the mapping (\ref{eq1_new}), expressed as $I_{n+1}=I_n+\epsilon \sin(\theta_n)$, the control parameter $\epsilon$ plays a key role. When $\epsilon=0$, notice that $I_{n+1}=I_n$, which means it is independent of time. Now, here is the pivotal point. When $\epsilon\ne0$, the algebraic form of the equation is disrupted, this disturbance is not just a numerical change -- it marks a profound break of symmetry, an algebraic break of symmetry.

\subsection{Order parameter and elementary excitations}

Let us begin by introducing an observable that meets the requirements similar to those of a typical order parameter at a continuous phase transition \cite{b2,sethna}. As we explored earlier, when $\epsilon=0$, the dynamics is regular, but as soon as $\epsilon\ne0$, chaos can emerge, leading to chaotic diffusion. However, chaotic diffusion is constrained due to the presence of two invariant spanning curves, one from the positive side and the other from the opposing side.

Given the symmetry of the phase space and considering we are dealing with diffusion and chaotic dynamics, the average action is not an ideal variable. Instead, a more suitable candidate is the root mean square of the squared action. Its value, when observed over a sufficiently long time, indicates the saturation of chaotic diffusion and is denoted as $I_{\text{sat}} \propto \epsilon^{\alpha}$. This variable aligns well with the characteristics of an order parameter. As $\epsilon$ approaches zero, it smoothly and continuously tends to zero, marking an ordered phase, and diverges from zero, indicating a chaotic phase.

A quick comparison with a transition in a ferromagnetic system can provide insight \cite{b2}. In such a system composed of interacting spins aligning with an external field, spontaneous magnetization ($m$) serves as the order parameter. Only local interactions define the magnetization at a null external field, dependent on the external temperature ($T$). Non-null magnetization is observed for temperatures below a critical point ($T_c$). However, once the temperature surpasses $T_c$, the ordered phase, characterized by aligned spins, breaks down, and null magnetization is observed. As $T$ approaches $T_c$ from below, the magnetization smoothly and continuously decreases to zero. The response of the order parameter to the external field gives the magnetic susceptibility ($\chi$), which diverges in this limit. These features align with the elements of a continuous phase transition.

\begin{figure}[t]
\centerline{\includegraphics[width=1.0\linewidth]{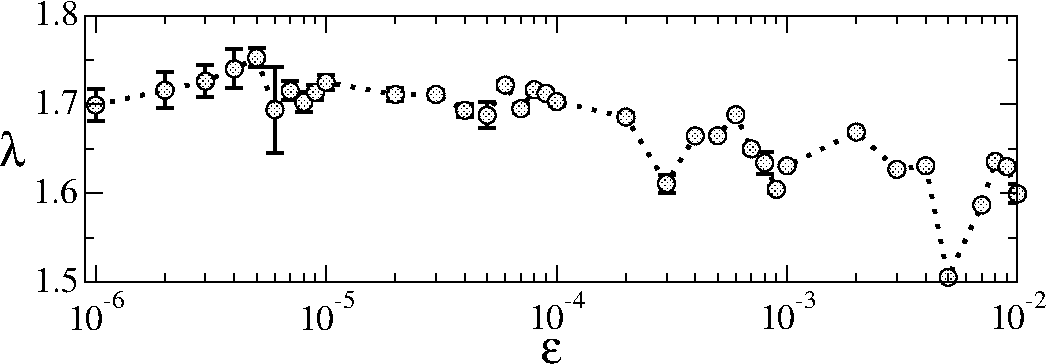}}
\caption{Plot of the positive Lyapunov exponent for a large range of control parameters $\epsilon\in[10^{-6},10^{-2}]$.}
\label{Fig2}
\end{figure}
In the chaotic model, once the control parameter $\epsilon$ is set apart from zero, the chaotic sea is born with a limited size, as discussed in Ref. \cite{ref1}. Figure \ref{Fig2} shows the positive Lyapunov exponent \cite{ref3} for a wide range of the control parameter $\epsilon\in[10^{-6},10^{-2}]$.

What catches our attention is the behavior of the positive Lyapunov exponent $\lambda$. It exhibits minimal variation, typically falling within $\lambda\in[1.5,1.75]$. This is quite striking when we consider a substantial range covered by the control parameter $\epsilon\in[10^{-6},10^{-2}]$, therefore marking four orders of magnitude.

This observation leads us to an interesting assumption: the chaotic sea has a size, and the chaotic dynamics possess a finite positive Lyapunov exponent. The nearly constant value of $\lambda$ is closely tied to the scaling invariance of the chaotic sea concerning the control parameter $\epsilon$.

\begin{figure}[t]
\centerline{\includegraphics[width=1.0\linewidth]{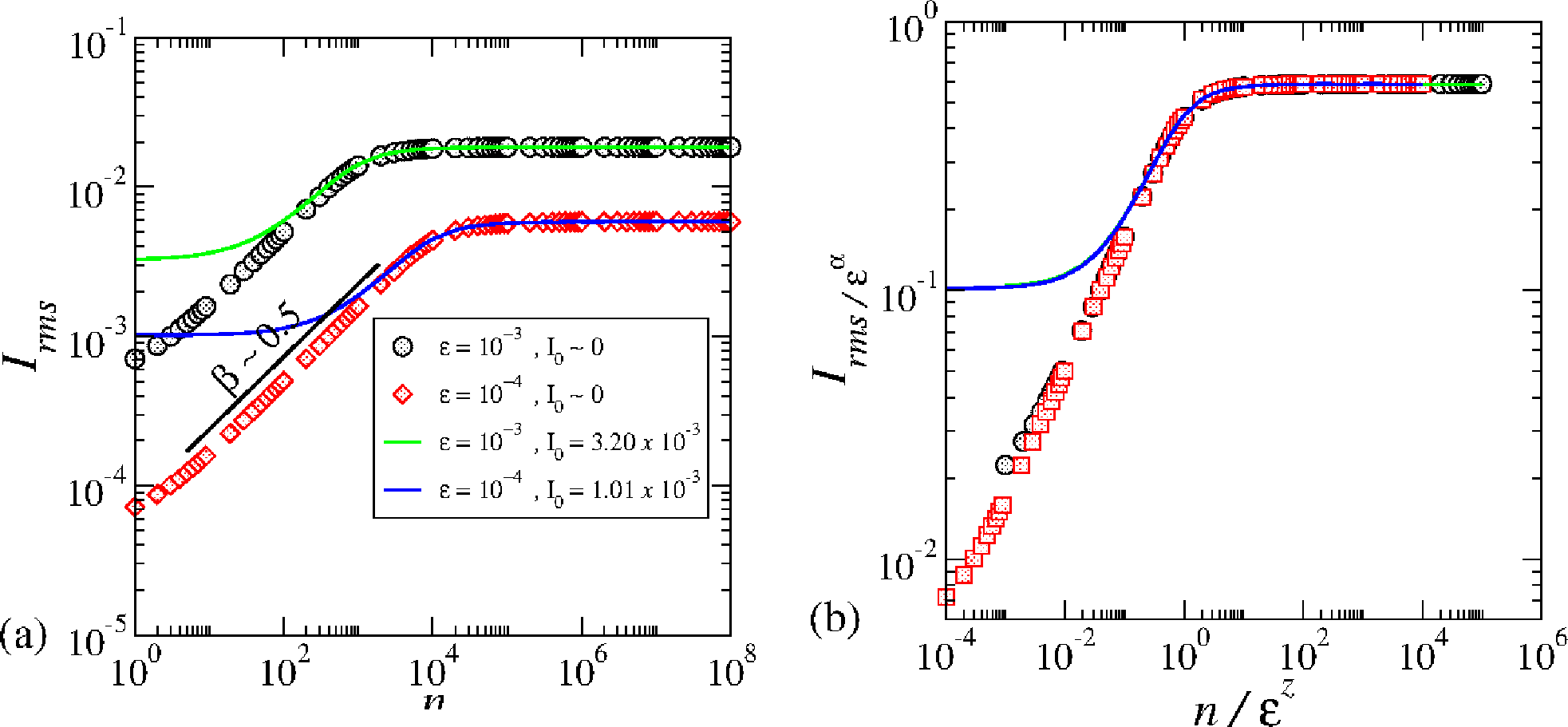}}
\caption{(a) Plot of different curves of $I_{rms}~vs.~n$ for the control parameters and initial action as labeled in the figure. (b) Overlap of the curves shown in (a) onto a single and universal plot.}
\label{Fig3}
\end{figure}

Now, let us go into the natural observable along the chaotic sea that serves as evidence for diffusion: the square root of the averaged squared action, defined as
\begin{equation}
I_{rms}=\sqrt{{{1}\over{M}}\sum_{i=1}^M{{1}\over{n}}\sum_{j=1}^nI^2_{i,j}}.
\label{eq2}
\end{equation}
Here, $M$ corresponds to an ensemble of different initial conditions, and $n$ is the length of time. In Figure \ref{Fig3}(a), the behavior of $I_{rms}$ unfolds as follows: for an initial action around $I_0\cong 0$, the curves of $I_{rms}\propto (n\epsilon^2)^{\beta}$ emerge, with the exponent $\beta\cong 1/2$, signifying particle diffusion akin to normal diffusion.

The term $\epsilon^2$ in the equation may seem arbitrary but is rooted in the dynamics. In Reference \cite{ref4}, this term was introduced to validate scaling assumptions. However, it can be derived analytically from the first mapping equation (\ref{eq1_new}). The nonlinear term $\sin(\theta_n)$ defines the elementary excitation of the dynamics. For chaotic dynamics and assuming statistical independence of the dynamical variables $\theta$ and $I$, and for small values of $I$, the first equation of mapping (\ref{eq1_new}) leads to an equivalent random walk dynamics with an average size of $\epsilon/\sqrt{2}$. This size becomes the elementary excitation of the system. Taking the square of the first equation of mapping (\ref{eq1_new}), averaging over an ensemble of different initial phases $\theta_0\in[0,2\pi]$, and assuming statistical independence between $I$ and $\theta$, we obtain $\overline{I^2}_{n+1}=\overline{I^2}_n+{{\epsilon^2}\over{2}}$. This equation allows us to derive the diffusion coefficient as $D={{\epsilon^2}\over{4}}$. A transformation of the difference equation into a differential equation yields the result $\overline{I^2}(n)=\overline{I^2}_0+n{{\epsilon^2}\over{2}}$, thereby analytically confirming the presence of the term $\epsilon^2$. As a short notice, this term has been introduced ad-hoc before and appeared naturally from the procedure.

As time evolves and with the presence of invariant spanning curves, we observe a fascinating behavior in the curve of $I_{rms,sat}\propto \epsilon^{\alpha}$ with $\alpha={{1}\over{1+\gamma}}$. This signifies a crucial transition in the system. The specific moment when the growth transitions to saturation is defined by $n_x\propto \epsilon^z$ where $z=-{{2\gamma}\over{\gamma+1}}$. The beauty of this observation lies in the curves overlapping harmoniously after a carefully applied scaling transformation, as illustrated in Figure \ref{Fig3}(b).
The overlapping of the curves shown in Figure \ref{Fig3}(b) serves as confirmation for a scaling invariance observed in chaotic dynamics near the transition from integrability to non-integrability for the mapping (\ref{eq1_new}). Notably, this scaling persists even when the initial action is small, as demonstrated by the continuous curves. These continuous curves were obtained through the analytical solution of the diffusion equation under specific boundary conditions \cite{celia}.

We notice that in the $\epsilon\rightarrow 0$ limit, the order parameter $I_{sat}\propto\epsilon^{\alpha}$ approaches zero continuously. The theory of second order phase transition \cite{cardy,leo} says that the susceptibility, that is, the response of the order parameter to the corresponding parameter $\epsilon$, must diverge in the above limit. The susceptibility is calculated as
\begin{equation}
\chi={{\partial I_{sat}}\over{\partial \epsilon}}=\left[{{1}\over{1+\gamma}}\right]{{1}\over{\epsilon^{{{\gamma}\over{1+\gamma}}}}}.
\end{equation}
Since $\gamma$ is a nonnegative number, in the limit of $\epsilon\rightarrow 0$, then $\chi\rightarrow\infty$, this is a clear signature of a second-order phase transition.

\subsection{Topological deffects}

We now go into the concept of topological defects. This terminology is borrowed from statistical mechanics, denoting the factors leading to the break of ergodicity in the dynamics. If the system's dynamics were chaotic, devoid of periodic points, and the average over a microcanonical ensemble equaled the time average, the system would be deemed ergodic. However, this is not the case due to periodic islands in the mixed phase space. These islands are akin to topological defects that disrupt the ergodic nature of the system. As the dynamics traverse close to these islands, a phenomenon known as stickiness is observed, subsequently altering the probability \cite{celia} of a particle to either survive or escape from a given region (for more details, refer to Ref. \cite{ref4}).

\section{Discussions}

Let us briefly review the chronological evolution of the idea conceived in Lancaster. Investigate some scaling properties observed in chaotic diffusion in systems described by nonlinear mappings. One characteristic is that two-dimensional and area-preserving mappings describe the dynamics; hence, the Liouville theorem is preserved. The mapping is defined in terms of two dynamical variables, one representing the action and the other the angle, always modulated $2\pi$. The characteristic of the mapping is that the angle diverges in the limit of vanishing action. It, therefore, leads the dynamics to be chaotic for small actions and shows a regular phase space for extensive enough. In the middle, there is a mixed type of dynamics with either chaos, periodic islands, or invariant tori. The interesting point was investigating the phase transition and the universality class from integrability to non-integrability. An exciting result was a scaling for the chaotic diffusion, which appeared in several systems and was characterized by critical exponents.

Physics evolves as we understand the problems and develop tools to characterize them. We started with scaling, which was investigated using a phenomenological approach. The procedure led to a homogeneous and generalized function allied with a set of three critical exponents connected by a scaling law. The procedure was validated in different chaotic systems.

In the second stage, after validating the phenomenological procedure for different models, we evolved to a more robust method, which led us to determine the position of the invariant tori and transform the equations of differences of the mapping into an ordinary differential equation. With this transformation, the time evolution of the average squared diffusion could be made analytically, which recovers the numerical simulations remarkably well.

As a last approach, the next step is to follow a more robust procedure seeking the solution of the diffusion equation by imposing specific boundaries and initial conditions. The diffusion equations give us the probability density to observe a certain particle with a specific action at a determined time. Knowledge of it is equivalent in statistical mechanics to the partition function, where all possible states can be extracted from there. Therefore, all momenta of the distribution can be obtained, and the critical exponents appear analytically. 

To finally determine the type of phase transition, we delved into four questions: identifying the broken symmetry, the topological defects, the elementary excitation, and proposing an order parameter. Since the scaling invariance is present and the characteristic of a second-order phase transition can be extracted from the order parameter (approaching zero) and its susceptibility (diverging) at the transition, we concluded that the transition from integrability to non-integrability shares the characteristics of a second-order phase transition.

The sequence of different steps and procedures led us to publish four books, two in Portuguese language \cite{book3,book4} by Blucher and two in English \cite{book1,book2} by Springer, hence marking the positive impact of the topic born in Lancaster earlier 2003.

\section{Final remarks}

In summary, the idea initially discussed in Peter's group is based on a solid research area. We developed an approach to investigate phase transition in chaotic systems using (i) a phenomenological approach with a set of critical exponents and scaling law (ii) a semi-analytical investigation where the discrete equation of the mapping is transformed into a differential equation and solved analytically as well as identifying the size of the chaotic sea by localization of the invariant tori (invariant spanning curve); (iii) by the solution of the diffusion equation giving the probability density to observe a particular particle with a given action at a specific time. 

Our results then conducted us to a transition from integrability to non-integrability by characterizing the basic elements that can be used to identify and classify a second-order phase transition in a dynamical system. We saw that the scaling present in the chaotic diffusion is linked to the limit size of the 
chaotic domain, leading to a set of critical exponents used to transform the curves of chaotic diffusion onto a universal curve. The order parameter was identified as $I_{sat}\propto\epsilon^\alpha$ with $\alpha={{1}\over{1+\gamma}}$, $\gamma>0$, where $\epsilon$ is the control parameter and that $I_{sat}\rightarrow 0$ when $\epsilon\rightarrow 0$. The susceptibility $\chi={{1}\over{1+\gamma}}{{1}\over{\epsilon^{{{\gamma}\over{1+\gamma}}}}}$ diverges in the limit of $\epsilon\rightarrow 0$. These two results recover well the case of the Fermi-Ulam model and the periodically corrugated waveguide by fixing $\gamma=1$. Moreover, such results are signatures of continuous phase transitions. 

We go beyond when discussing the elementary excitations produced by the nonlinear function, leading the dynamics at the low action domain to behave as a random walk particle. The existence of the periodic islands was interpreted as topological defects in the phase space modifying the system's transport properties, leading to sticky dynamics. The discussion presented here allows us to conclude that the phase transition from integrability to non-integrability is analogous to a second-order phase transition.

We must say the results discussed and the formalism used can be extended to many other different types of phase transitions in dynamical systems, including a transition from limited to unlimited chaotic diffusion \cite{antonio} and also from limited to unlimited Fermi acceleration \cite{EDL1,EDL2} in time-dependent billiard systems \cite{lenz}.

E.D.L. acknowledges support from Brazilian agencies CNPq (No. 301318/2019-0, 304398/2023-3) and FAPESP (No. 2019/14038-6 and No. 2021/09519-5).


\begin{thebibliography}{9}

\bibitem{barabasi} A. -L. Barab\'asi, H. E. Stanley, {\it Fractal Concepts in Surface Growth} (Cambridge University Press, Cambridge, 1995).

\bibitem{edl1}
E. D. Leonel, P. V. E. McClintock. Chaotic properties of a time-modulated barrier. {\em Phys. Rev. E} {\bf 2004}, {\em 70}, 016214(1)--016214(11).

\bibitem{edl2}
E. D. Leonel, P. V. E. McClintock. Dynamical properties of a particle in a time-dependent double-well potential. {\em J. Phys. A} {\bf 2004}, {\em 37}, 8949--8968.

\bibitem{edl3}
E. D. Leonel, P. V. E. McClintock. A hybrid Fermi-Ulam-bouncer model. {\em J. Phys. A} {\bf 2005}, {\em 38}, 823--839.

\bibitem{edl4}
E. D. Leonel, P. V. E. McClintock. A crisis in the dissipative Fermi accelerator model. {\em J. Phys. A} {\bf 2005}, {\em 38}, L425--L430.

\bibitem{edl5}
E. D. Leonel, P. V. E. McClintock. Scaling properties for a classical particle in a time-dependent potential well. {\em Chaos} {\bf 2005}, {\em 15}, 033701(1)--033701(7).

\bibitem{edl6}
E. D. Leonel, P. V. E. McClintock. Dissipative area-preserving one-dimensional Fermi accelerator model. {\em Phys. Rev. E} {\bf 2006}, {\em 73}, 066223(1)--066223(5).

\bibitem{edl7}
E. D. Leonel, P. V. E. McClintock. Effect of a frictional force on the Fermi-Ulam model. {\em J. Phys. A} {\bf 2006}, {\em 39}, 11399--11415.

\bibitem{edl_prl}
E. D. Leonel, P. V. E. McClintock, J. K. L Silva. Fermi-Ulam Accelerator Model under Scaling Analysis. {\em Phys. Rev. Lett.} {\bf 2004}, {\em 93}, 014101(1)--014101(4).

\bibitem{marion} M. McClintock, Quest for Innovation: History of the First Ten Years of Lancaster University {\bf 1974}, Construction Press. 

\bibitem{lich} A. J. Lichtenberg, M.A. Lieberman. {\it Regular and Chaotic Dynamics}(Appl. Math. Sci. {\bf 38}, Springer Verlag, New York, 1992).

\bibitem{fermi} E. Fermi, Phys. Rev. {\bf 75}, 1169 (1949).

\bibitem{joelson1} J. D. V. Hermes, M. A. dos Reis, I. L. Caldas, E. D. Leonel. {\em Chaos, Solitons and Fractals}, {\bf 2022}, {\em 162}, 112410

\bibitem{joelson2} Y. H. Huggler, J. D. V. Hermes, E. D. Leonel. {\em Chaos: an Interdisciplinary Journal of Nonlinear Science}, {\bf 2022}, {\em 32}, 093125.

\bibitem{celso} C. V. Abud, I. L. Caldas. {\em Physica D}, {\bf 2015}, {\em 308}, 34.

\bibitem{leonel_waveguide}  E. D. Leonel, Corrugated waveguide under scaling investigation. {\em Phys. Rev. Lett.}, {\bf 2007}, 98, 114102.


\bibitem{pub1} E. D. Leonel, J. A. de Oliveira, F. Saif. {\em J. Phys. A: Math. Theor.},  {\bf 2011}, {\bf 44}, 302001

\bibitem{pub2} E. D. Leonel, J. Penalva, R. M. N. Teixeira, R. N. Costa Filho, M. R. Silva, J. A. de Oliveira. {\em Phys. Lett. A}, {\bf 2015}, {\em 379}, 1808.

\bibitem{pub3} E. D. Leonel, C. M. Kuwana, M. Yoshida, J. A. de Oliveira. {\em EPL}, {\bf 2020}, {\em 131}, 10004

\bibitem{pub4} E. D. Leonel, M. Yoshida, J. A. de Oliveira. {\em EPL}, {\em EPL}, {\bf 2020}, 20002.

\bibitem{book1} E. D. Leonel. {\em Scaling Laws in Dynamical Systems}, {\bf 2021}, Sringer.

\bibitem{book2} E. D. Leonel. {\em Dynamical Phase Transitions in Chaotic Systems}, {\bf 2023}, Springer.

\bibitem{b2} R. K. Patria, Statistical Mechanics, Elsevier (2008)

\bibitem{b3} F. Reif, Fundamentals of statistical and thermal
physics, New York: McGraw-Hill, (1965)

\bibitem{b4} V. Balakrishnan, Elements of nonequilibrium statistical
mechanics, Ane Books India, New Delhi (2008)

\bibitem{b5}  G. J. Sussman, J. Wisdom, and M. E. Mayer, Structure and
Interpretation of Classical Mechanics, MIT Press, Cambridge (2001)

\bibitem{b6} S. A. Holgate, Understanding Solid State Physics, CRC Press, Boca 
Ratom, Florida-USA (2010).

\bibitem{ref2} E. G. Altmann, J. S. E. Portela, T. T\'el, Rev. Mod.
Phys. {\bf 85} 869 (2013).

\bibitem{sethna} J. P. Sethna. {\em Entropy, Order Parameters, and Complexity}, {\bf 2006}. Oxford, Oxford University Press.

\bibitem{ref1} E. D. Leonel, J. A. de Oliveira, F. Saif, J. Phys. A, 
{\bf 44}, 302001 (2011)

\bibitem{ref3} J. P. Eckmann, D. Ruelle, Rev. Mod. Phys., {\bf 57} 617 (1985).

\bibitem{ref4} E. D. Leonel, J. Penalva, R. M. N. Teixeira, R. N. Costa Filho, 
M. R. Silva, J. A. de Oliveira, Phys. Lett. A, {\bf 379}, 1808 (2015).

\bibitem{celia} E. D. Leonel, C. M. Kuwana, J. Stat. Phys. {\bf 170}, 69 (2018)

\bibitem{cardy} J. Cardy, Scaling and Renormalization in Statistical Physics. Cambridge University Press,  1996.

\bibitem{leo} L. P. Kadanoff, Statistical Physics: Statics, Dynamics and Renormalization. World Scientific, Singapore, 1999.

\bibitem{book3} E. D. Leonel. {\em Fundamentos da F\'isica Estat\'istica}, {\bf 2015}, S\~ao Paulo: Blucher.

\bibitem{book4} E. D. Leonel. {\rm Invari\^ancia de Escala em Sistemas Din\^amicos N\~ao Lineares}, {\bf 2019},  S\~ao Paulo, Blucher.

\bibitem{antonio} R. Aguilar-Sanchez, E. D. Leonel, and J. A. Mendez-Bermudez. Phys. Lett. A {\bf 377}, 3216 (2013).

\bibitem{EDL1} E. D. Leonel, L. A. Bunimovich, Phys. Rev. Lett. {\bf 104}, 224101 (2010).

\bibitem{EDL2} D. F. M. Oliveira, E. D. Leonel, Phys. Lett. A {\bf 374}, 3016 (2010).

\bibitem{lenz} F. Lenz, F. K. Diakonos, and P. Schmelcher, Phys. Rev. Lett. {\bf 100}, 014103 (2008).

\end{thebibliography}
\end{document}